\documentclass[sigconf]{acmart}
\AtBeginDocument{%
  \providecommand\BibTeX{{%
    \normalfont B\kern-0.5em{\scshape i\kern-0.25em b}\kern-0.8em\TeX}}}

\setcopyright{acmcopyright}
\copyrightyear{2023}
\acmYear{2023}
\usepackage{multirow}
\acmConference[In submission to CIKM’23 workshop]{}{}{}

\acmPrice{15.00}

\begin{document}

\title{Multiple Key-value Strategy in Recommendation Systems
Incorporating Large Language Model}


\author{Dui Wang}

\email{wangdui@meituan.com}
\affiliation{%
  \institution{Meituan Inc}
  \country{Beijing, China}
}

\author{Xiangyu Hou}

\email{houxiangyu02@meituan.com}
\affiliation{%
  \institution{Meituan Inc}
  \country{Beijing, China}
}

\author{Xiaohui Yang}

\email{yangxiaohui04@meituan.com}
\affiliation{%
  \institution{Meituan Inc}
  \country{Beijing, China}
}

\author{Bo zhang}

\email{zhangbo58@meituan.com}
\affiliation{%
  \institution{Meituan Inc}
  \country{Beijing, China}
}

\author{Renbing Chen}

\email{chenrenbing@meituan.com}
\affiliation{%
  \institution{Meituan Inc}
  \country{Beijing, China}
}

\author{Daiyue Xue}

\email{xuedaiyue@meituan.com}
\affiliation{%
  \institution{Meituan Inc}
  \country{Beijing, China}
}


\begin{abstract}
  Recommendation system (RS) plays significant roles in matching users’ information needs for Internet applications, and it usually utilizes the vanilla neural network as the backbone to handle embedding details. Recently, the large language model (LLM) has exhibited emergent abilities and achieved great breakthroughs both in the CV and NLP communities. Thus, it is logical to incorporate RS with LLM better, which has become an emerging research direction. Although some existing works have made their contributions to this issue, they mainly consider the single key situation (e.g. historical interactions), especially in sequential recommendation. The situation of multiple key-value data is simply neglected. This significant scenario is mainstream in real practical applications, where the information of users (e.g. age, occupation, etc) and items (e.g. title, category, etc) has more than one key. Therefore, we aim to implement sequential recommendations based on multiple key-value data by incorporating RS with LLM. In particular, we instruct tuning a prevalent open-source LLM (Llama 7B) in order to inject domain knowledge of RS into the pre-trained LLM. Since we adopt multiple key-value strategies, LLM is hard to learn well among these keys. Thus the general and innovative shuffle and mask strategies, as an innovative manner of data argument, are designed. To demonstrate the effectiveness of our approach, extensive experiments are conducted on the popular and suitable dataset MovieLens which contains multiple keys-value. The experimental results demonstrate that our approach can nicely and effectively complete this challenging issue.
\end{abstract}




\maketitle

\section{Introduction}
Recommendation Systems (RS) aim to address the issue of online information overload and generate target items for the user according to the user's preferences, which are often achieved by their historical interactions. In the literature on recommendation systems, most of the existing works\cite{kang2018self,sun2019bert4rec,guo2017deepfm} train deep neural networks to extract user and item features, enabling more complex and challenging recommendation applications, such as playlist generators~\cite{kowald2020unfairness,singh2022novel} for listeners, product recommendations~\cite{sun2022revisiting,liu2022ecommerce,xie2022decoupled} for customers, and content recommendations~\cite{mittal2020smart,perez2021content,xie2023rethinking,volkovs2017content} for readers, among others.

On the other hand, Large language model (LLM)~\cite{touvron2023llama,radford2019language,scao2022bloom,touvron2023llama2} has sparked a revolution in the field of natural language processing (NLP). With the rapid increase in LLM parameters, LLM has shown impressive emergent abilities (e.g. reasoning~\cite{zhang2023multimodal}, understanding~\cite{lian2023llm}, in-context few-shot learning~\cite{dang2022prompt}). Given the success of LLMs, RS may benefit from their prosperity by incorporating them~\cite{lin2023can}. However, the key challenge of this issue is how to convert structured data to natural language. The raw data of recommendation systems are discrete and key-based, such as id, age, occupation for users, and title, category, actions for items. In contrast, the input of LLM should be continuous natural language text, which contains abundant semantic information. To address this challenging problem, various  works have been conducted and achieved promising results. For example,~\cite{zhang2023recommendation} proposed various manual-designed templates to generate instruction data, converting the discrete and structured raw data to natural language text format. Others~\cite{liu2023chatgpt,hou2023large} explored the ability of LLM to understand recommendation tasks using zero-shot and few-shot strategies. Additionally, ~\cite{geng2022recommendation} proposed a flexible and unified text-to-text paradigm, which unified various recommendation tasks in a shared framework. 

Although some existing works have made their contributions to successfully incorporating RS with LLM, they mainly consider the single key (e.g. historical interactions) scenario in the recommendation task. Specifically, the raw data of RS are often discrete and key-based. For example, the MovieLens dataset, as the popular benchmark for RS, has various keys for both users(e.g. gender, age, occupation) and items(e.g. title, rating, category). In practical applications, it is common to utilize various key information about users and items to improve performance, instead of just utilizing a single key. Therefore, utilizing multiple key-value data is necessary and effective for RS, along with the combination of LLM. However, it is difficult to let an LLM adapt to recommendation systems, as LLM may result in unsatisfactory performance due to the increase in model inputs and semantic complexity. Therefore, it is necessary and worthwhile to invest in this issue and provide valuable insights to explore the potential of LLM in RS.

To bridge this research gap, we take into account multiple key-value data and utilize LLM to complete sequential RS. To the best of our knowledge, this is the first principled work for sequential RS on multiple key-value data by utilizing LLM. In this section, we first analyze the core challenge, which is the mismatch problem between the structured raw data and the desired natural language text data. To address this issue, we can conclude into two methodologies: 1) Adapting LLM to the structured raw data by modifying the construction of LLM; 2) Adapting the structured raw data to LLM by converting these data to continuous natural language text data. While the first solution can better handle the key-value data, modifying the model construction or adding the components may disturb the model structure and result in knowledge loss. Regarding the second solution, it is more flexible and only requires an extra process of training data without changing the model. Furthermore, this strategy can effectively preserve knowledge of the pre-train model. Based on the above analysis, we adopt the second strategy and propose a template that converts the structured raw data to natural language text data. However, multiple key-value data may lead to poor learning performance due to the large number of input tokens. To address this problem, we further propose two novel data argument strategies to enhance the semantic association between these keys and candidates. Specifically, we propose a shuffle strategy to shuffle key-value pairs or candidates and a mask strategy to randomly mask out part of key values. These augmented data are associated with the same label. In addition, it is possible to flexibly combine these proposed strategies based on specific situations. We conduct extensive experiments on the popular dataset (MovieLens), and the results demonstrate the effectiveness of the proposed strategies.

\section{Notations and Preliminaries}
In this work, we discuss a typical sequential recommendation for RS. Let $I=\{i_{1},i_{2},...,i_{|I|}\}$ be the set of items and $U=\{U_{1},U_{2},...,U_{|U|}\}$ be the set of users. We define $K^{u}=\{k^{u}_{1},k^{u}_{2},...,k^{u}_{|K^{u}|}\}$ as the set for users, while $K^{i}=\{k^{i}_{1},k^{i}_{2},...,k^{i}_{|K^{i}|}\}$ as the set of keys for items. $V^{u}$ and $V^{i}$ are the corresponding value set of keys for users and items, respectively. Regarding both keys and values, we can define $KV^{u}_{j}=\{<k^{u}_{1},V^{j}_{1}>,<k^{u}_{2},V^{j}_{2}>,...,<k^{u}_{|K^{u}|},V^{j}_{|K^{u}|}>\}$, $j\in U$ as key-value pairs for the $j$-index user, and $KV^{i}_{j}$ is define for the $j$-index item in the same way. Due to the sequential recommendation, one of $K^{i}$ is the historical interaction. We construct a user's action sequence $S^u=\{S^U_{1},S^U_{2},..,S^U_{|S^U|}\}$. The LLM can be defined as $F: S^u \rightarrow R $, where $R$ is the re-ranked sequence of the candidate set and the first item in the line is the target user's next item. As we apply LLM, we define $Ins=\{x_{1},x_{2},...,x_{m}\}$, where $Ins$ represents a set of $m$ instruction samples to be constructed for LLM training. 
\section{Methodology}
In this section, we present our methodology in the following sections. Specifically, to convert structured multiple key-value data to natural language format, we propose a template designed for large language model input structure. Although the core challenges mentioned above can be alleviated, the model may still perform poorly due to the increased volume and complexity of input data. Therefore, we also propose shuffle and mask strategies to enhance the semantic association between multiple key values and candidate sets.
\subsection{Instruction Format}
Considering the sequential recommendation task involving multiple key-value pairs, we first construct the following template to facilitate the conversion of the structured data into natural language format.
\begin{figure}[h]
  \centering
  \includegraphics[width=\linewidth]{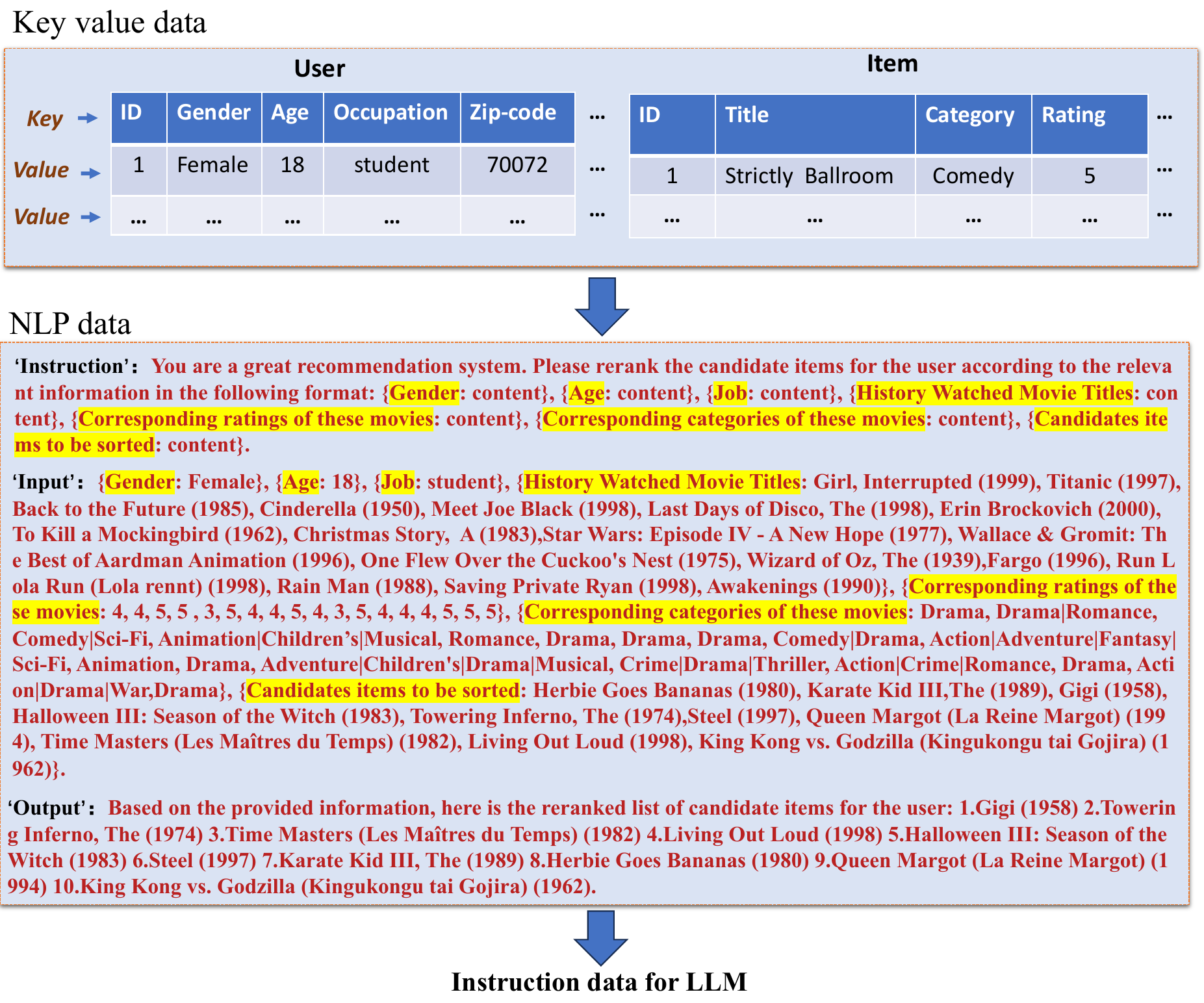}
  \caption{Example instructions where keys are in yellow background. We follow the classical format of LLM instructions, which includes 'Instruction', 'Input', and 'Output'.}
 \label{instruction_format}
\end{figure}

As shown in Figure~\ref{instruction_format}, the raw data for RS is constructed in a key-value format. The proposed template can convert the raw data into text data that satisfies LLM input requirements. Concretely, we consider user keys (e.g. gender, occupation, age, historical interactions) and item keys (e.g. title, category, rating) when constructing the training data. For sequential recommendations, we follow the previous work~\cite{rendle2010factorizing,he2017translation,he2016fusing,kang2018self} to generate a randomized candidate set, which contains one positive sample and several randomly chosen negative samples. Additionally, the first item of the output is the positive sample (the next item), while the others are all negative samples from the candidate set. 

\subsection{Adopting Shuffle Strategy}
\begin{figure}[h]
  \centering
  \includegraphics[width=\linewidth]{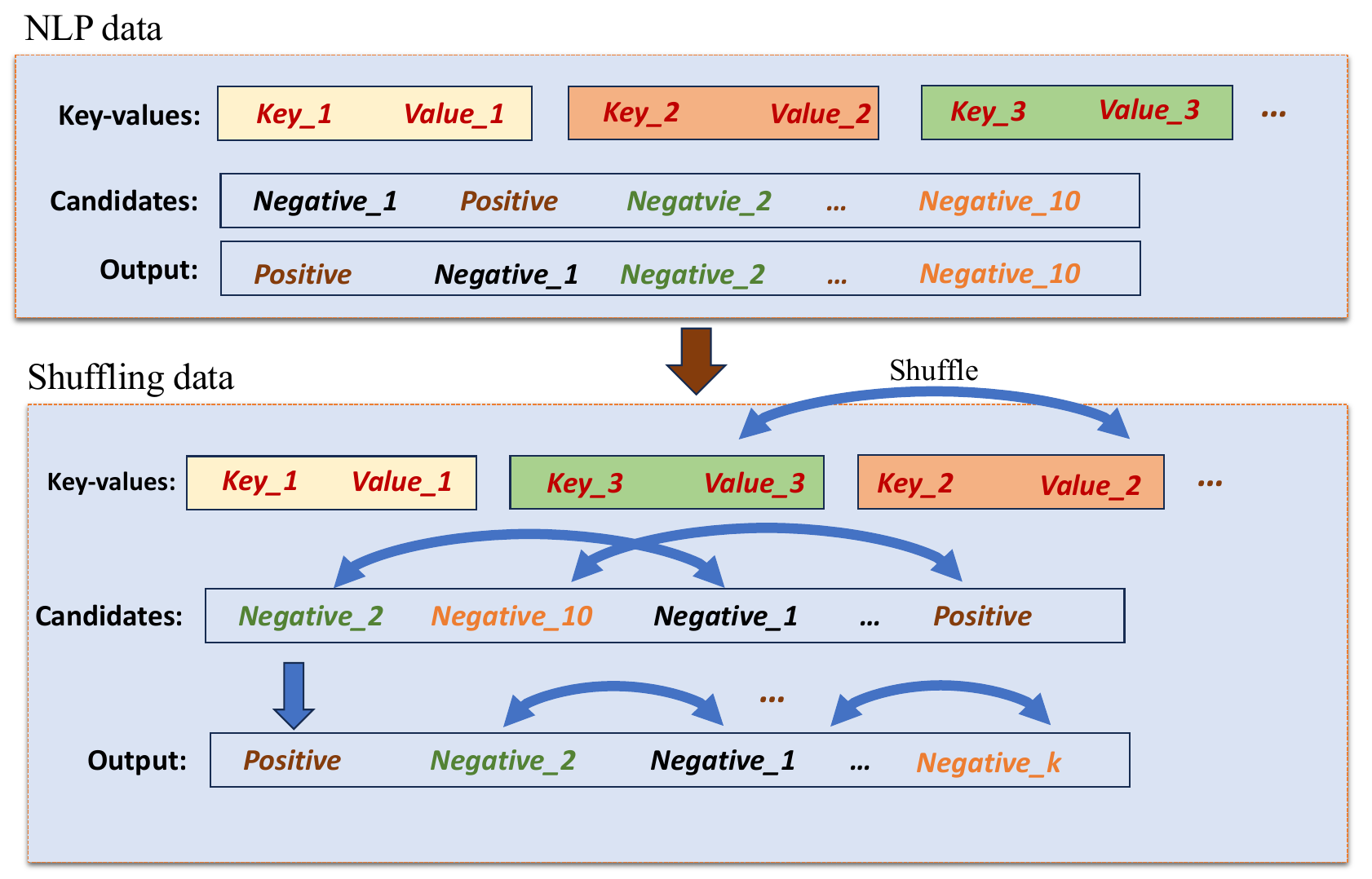}
  \caption{The proposed shuffle strategy. We shuffle key-value pairs or other lists in the instructions. The crossed arrows indicate shuffling the list, rather than exchanging specific items.}
 \label{shuffle_strategy}
\end{figure}

Although the proposed template can alleviate the core challenge, it also increases the data volume and complexity of LLM's input, which can lead to poor performance. To address this problem, we apply the data argument strategy to enhance the association between these keys and candidates. Firstly, this section presents the shuffle strategy, which is shown in Figure ~\ref{shuffle_strategy}. 

Empirically, the order of key-value pairs may not influence the resulting output. From a human perspective, the key-value pairs $[<Gender>,<Occupation>]$ and $[<Occupation>,<Gender>]$ have identical semantic information and thus result in identical output. Motivated by this observation, we propose a shuffle strategy to randomly rearrange these key-value pairs or other information in list format, including the candidate list and output list.  
By adopting this strategy, we have the flexibility to decide on applied components. We can shuffle only a single component including key-value pairs, candidates set, and output set. Additionally, shuffling the combination of these components or both is also available. After data enhancement, both generated data and raw data are added to the training dataset. Given a finite number of training data, this strategy can well strengthen the attention of these components and alleviate the impact of pairs order on the results. 
\subsection{Adopting Mask Strategy}
\begin{figure}[h]
  \centering
  \includegraphics[width=\linewidth]{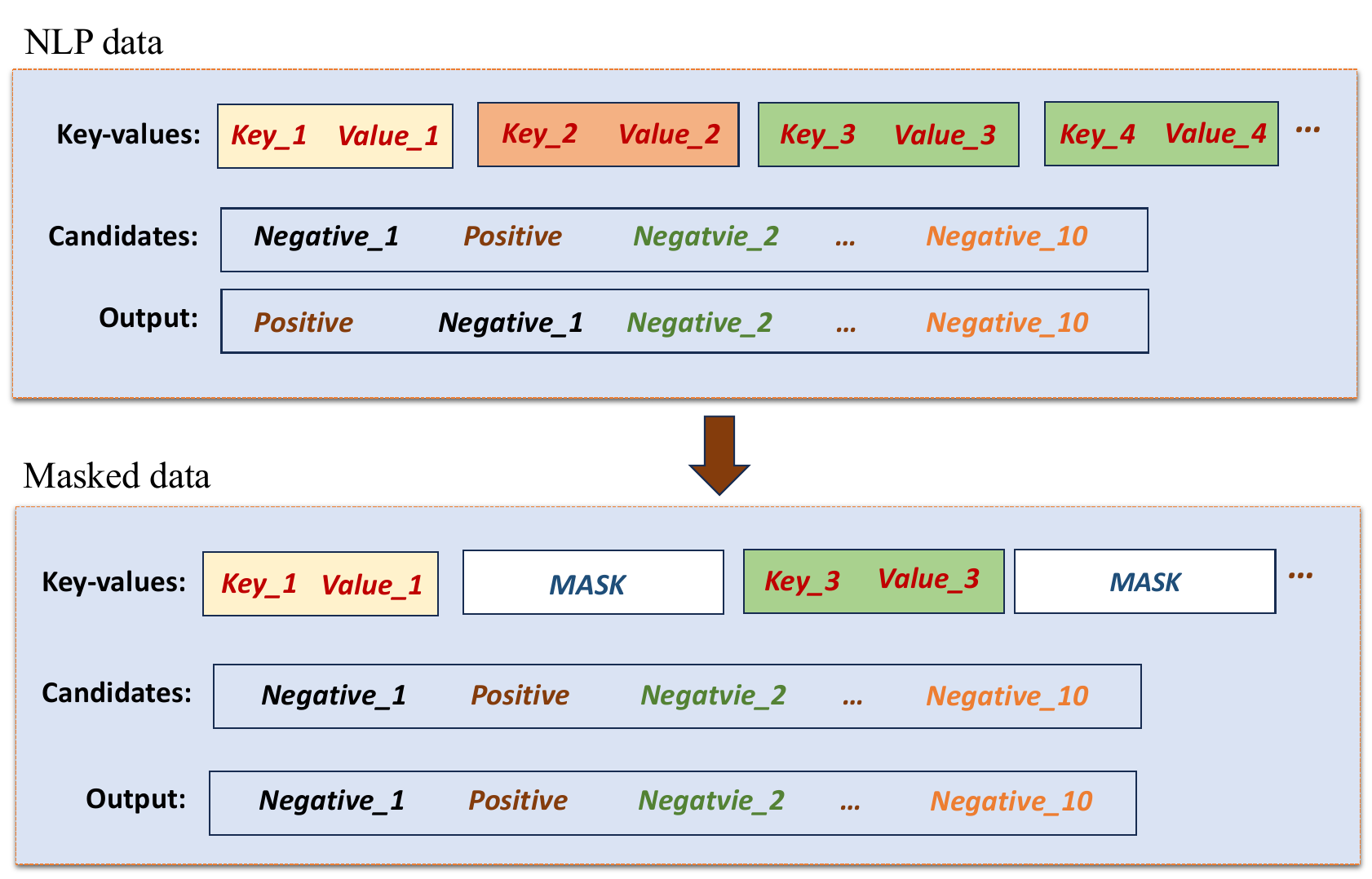}
  \caption{The proposed mask strategy aims to enhance the decision-making ability of prominent keys by masking out certain key-value pairs, as shown in the white background.}
 \label{mask_strategy}
\end{figure}

In addition to the shuffle strategy, we also propose a mask strategy. While the shuffle strategy aims to enhance the association between different components, it assumes that the order of these components influences the outcomes equally. However, different keys actually have different contributions to the outcomes, meaning that some keys may have higher importance weights then others. Empirically, some prominent keys, not all keys in some situations, can also result in true outcomes, such as historical interactions, category of items, or others. Motivated by this observation, we propose the mask strategy that masks out part of key-value pairs to strengthen the association between some prominent keys and the results, owing to the high importance weights of these keys. This strategy is shown in Figure~\ref{mask_strategy}.

Concretely, we can manually select which keys are retained and which ones to mask out. Empirically, the maintained keys are more critical for the results. However, in this paper, due to the huge cost of training LLM, we choose the simplest way, which randomly masks out some keys while always maintaining historical interactions. Differing from the shuffle strategy, this strategy has a hyper-parameter to determine the number of masked keys, and the setting of this parameter should be customized according to specific applications and the training dataset. By adopting this strategy, we can strengthen the association between prominent keys and results, thereby improving the learning performance.  

\section{Experiments}
\subsection{Setup}
\subsubsection{Datasets} Considering the specific setting, the \textbf{MovieLens} dataset best satisfies the multiple key-value format. Most of these values are in text format, rather than in encoded format. It is difficult to find other datasets that fulfill these conditions, as values of keys from other datasets are often in encoded format. We follow the same processing procedure from ~\cite{kang2018self}. Concretely, we discard users and items with fewer than $5$ related actions and use timestamps to determine the sequence order of actions. The keys for users are '\textit{userID}', \textit{gender'}, \textit{'age'}, \textit{'occupation'}, and \textit{'zip code'}. The keys for items are \textit{'itemID'}, \textit{'title'}, and \textit{'category'}. Additionally, we maintain the actions keys, including \textit{'history watched movie title'}, \textit{'corresponding rating of these movies'}, and 
 \textit{'corresponding category of these movies'}. By adopting our methods, we can construct 6040 instructions, one for each user, during both the training and testing processes. Moreover, we just randomly select $1000$ test samples due to the huge cost of the evaluation.

\subsubsection{Model}
We utilize the popular LLaMA-$7B$\cite{touvron2023llama} as the backbone and fine-tune its open-source pre-trained model. LLaMA is known for its effectiveness and competitiveness compared with other open-source LLMs. It has been widely used in numerous applications and has achieved significant success in supervised fine-tuning LLM. We first download the pre-trained foundation model of LLaMA-$7B$. Then we prepare instruction data using our proposed strategies and fine-tune LLM on this data, which includes both raw and enhanced data. 

\subsubsection{Configuration}
Regarding constructing instruction data, we discard \textit{'id'}, \textit{'zip code'}, and \textit{'timestamp'} of users due to its encoded format. Due to the contextual limit of input tokens, we truncate the generated behavioral sequence with a maximum of $20$ items, i.e., $|S^U|=20$. We generated a candidate set by randomly selecting $9$ negative samples and a positive sample (the next item) and utilize their titles to represent themselves. No extra keys exist for the candidate set.

\subsubsection{Baselines} 
To the best of our knowledge, this is the first work that studies the incorporation of RS with LLM to complete sequential recommendation tasks on multiple key-value data. Therefore, it is challenging to choose baselines that utilize both RS and LLM to handle sequential recommendation tasks. Thus we select the challenging GPT-4 to compare with our approach.

\subsubsection{Our method} 
Regarding our methods, we summarize various strategies as follows: 1) 
Multiple\_KVs. This approach utilizes the proposed template. We term it InstructMK; 2) KVs\_Shuffle. We shuffle the candidate list and the output list in the 'Input' and 'Output' components, respectively. We term these two methods InstructMK-SC and InstructMK-SO, respectively; 3)KVs\_Mask. We mask parts of key-value pairs in different degrees. Concretely, we mask out 3,4 of all 7 key-value pairs and term them InstructMK-M1 and InstructMK-M2, respectively. We also mask out 3 and 4 keys at the same time and term this method InstructMK-M3. 

All methods generate an equal number of enhanced data, with a ratio of original to enhanced data of 1:4. Specifically, InstructMK-M3 generates enhanced data with a ratio of 1:2 by masking out 3 keys, and the same applies to masking out 4 keys, resulting in a final ratio of 1:4 for InstructMK-M3. Additionally, all other combinations of these strategies (e.g. shuffle and mask) are feasible but need future verification in specific cases. In this work, we mainly want to provide methodological and directional value.

\subsection{Training and Evaluation}
To complete training and evaluation, we follow the previous work~\cite{kang2018self} by splitting the historical sequence $S^U$ into two parts:(1) the most recent action $S^U_{|S^U|}$ for testing; (2) the second most recent action $S^U_{|S^U|-1}$ for training. When preparing training data, the historical interactions sequence contains action $S^U_{|S^U|-2}$ and is in time order. We then chose the next item ($S^U_{|S^U|-1}$), which also exists in the candidate set, as the label of the corresponding instance. Similarly, we utilize action $S^U_{|S^U|}$ for the testing process.

To evaluate recommendation performance, we adopt two common Top-N metrics: Hit Rate$@$K, and NDCG$@$K. In this paper, we set $K$ as 1, 3, and 5. Since HR$@1$ is equal to NDCG$@1$, we only report HR$@1$. As the output of LLM is generative and may result in out-of-scope results, we also propose an Error Rate metric to evaluate the output of LLM. It can be defined as $ER=\frac{\sum^{|U|}_{j} I(F(j))} {|U|}$, where $I(.)$ is a indicator function. For the $j$-index sample, if every title in the outputs of LLM is correct, $I(F(j))=1$, and 0 otherwise.

\subsection{Results and Analysis}
\begin{table}[h]
  \label{tab:commands}
 \caption{Results (HR$@$K) on MovieLens.}
  \begin{tabular}{cccccl}
    \toprule
     {Strategy} &{Method}& {HR@1} &{HR@3} & {HR@5} \\
    \midrule
    GPT                                & GPT-4                & 0.3080 & 0.4260& 0.4820  \\
    \midrule
    Multiple\_KVs                        & InstructMK           & 0.6460 & 0.7670& 0.8270  \\
    \midrule
    \multirow{2}*{KVs\_Shuffle}   & InstructMK-SC        & 0.7130 & 0.8000& 0.8630  \\
       ~     & InstructMK-SO        & \textbf{0.7300} & \textbf{0.8070}& \textbf{0.8630}  \\
    \midrule
     \multirow{3}*{KVs\_Mask}       & InstructMK-M1        & 0.6980 & 0.7690& 0.8310  \\
          & InstructMK-M2        & 0.6790 & 0.7710& 0.8360  \\
          & InstructMK-M3        & 0.7020 & 0.7830& 0.8360  \\
    \bottomrule
  \end{tabular}
  \label{res_1}
\end{table}

\begin{table}[h]
  \label{tab:commands}
   \caption{Results (NDCG@K and ER) on MovieLens.}
  \begin{tabular}{cccccl}
    \toprule
     {Strategy} &{Method}& {NG@3} &{NG@5} & {ER} \\
    \midrule
    GPT                                & GPT-4                & 0.3777 & 0.4007 & 0.020  \\
    \midrule
    Multiple\_KVs                        & InstructMK           & 0.7165 & 0.7411 & 0.003  \\
    \midrule
    \multirow{2}*{KVs\_Shuffle}   & InstructMK-SC        & 0.7630 & 0.7886 & 0.006  \\
       ~                            & InstructMK-SO        & \textbf{0.7733} & \textbf{0.7962} & \textbf{0.000}  \\
    \midrule
     \multirow{3}*{KVs\_Mask}        & InstructMK-M1        & 0.7387 & 0.7649 & 0.007  \\
                                      & InstructMK-M2        & 0.7323 & 0.7591 & 0.009  \\
                                      & InstructMK-M3        & 0.7478 & 0.7697 & 0.011  \\
    \bottomrule
  \end{tabular}
  \label{res_2}
\end{table}

The experimental results are shown in the Tabel \ref{res_1} and \ref{res_2}. We can observe that: 1) The shuffle and mask strategies are all effective and can improve the results compared to the baseline GPT-4; 2) Among our proposed method, the shuffle strategy on the output set achieves the best performance; 3) Since we conducted experiments on a single dataset with a limit of 6 keys, the value of the mask strategy may be affected. However, our experiments show that this strategy is effective and results in significant improvements; 4) When it comes to the specific mask strategy, masking too many keys is not encouraged for model training. In our experiments, the combination of different mask degrees leads to better performance; 5) We find that our approach contributes more improvements to HR$@$1 and HR$@$3 for HR$@$K, indicating that our approach can increase the likelihood of the true next-item being ranked first; 6) Regarding baseline, GPT-4 lacks domain knowledge and cannot understand multiple key-value data well, resulting in poor performance.

\subsection{Limitations and Future}
Due to the input limit of LLM and the high training cost, we followed the approach of previous works and generated a behavioral sequence with a maximum length of 20 and a candidate set with a length of 10. Additionally, it was challenging to find suitable datasets that satisfy our setting, where the dataset should contain multiple key-value pairs and its values should be in natural language format. Given this unique situation and the significant cost of training large models mentioned above, we chose to use only the single MovieLens dataset. However, in the future, we plan to explore more suitable datasets and larger candidate sets.
\section{Conclusion}
This paper focuses on sequential recommendation scenarios based on multiple key-value data and incorporates RS with LLM. Specifically, we propose a conversion template to transform raw data into text data. Moreover, to address the challenge of increasing input number and complexity, we introduce two flexible and general data augmentation strategies: random shuffling of sets and masking of key-value pairs. Additionally, we propose the error rate as a metric to quantify the error rate of the model output.

\bibliographystyle{ACM-Reference-Format}
\bibliography{conf.bib}
\end{document}